\documentclass[fleqn,usenatbib]{mnras}

\usepackage{newtxtext,newtxmath}

\usepackage[T1]{fontenc}
\DeclareRobustCommand{\VAN}[3]{#2}
\let\VANthebibliography\thebibliography
\def\thebibliography{\DeclareRobustCommand{\VAN}[3]{##3}\VANthebibliography}

\usepackage{graphicx}	
\usepackage{amsmath}	

\title[APEX observations of NGC 1977 proplyds]{An APEX search for carbon emission from NGC 1977 proplyds}

\author[T. J. Haworth et al.]{Thomas J. Haworth$^{1}$\thanks{E-mail: t.haworth@qmul.ac.uk}, Jinyoung S. Kim$^{2,3}$,  Lin Qiao$^1$, Andrew J. Winter$^4$, Jonathan P. Williams$^5$
\newauthor
Cathie J. Clarke$^6$, James E. Owen$^7$, Stefano Facchini$^{8,9}$, Megan Ansdell$^{10}$, Mikhel Kama$^{11}$, Giulia Ballabio$^1$
\\
$^{1}$Astronomy Unit, School of Physics and Astronomy, Queen Mary University of London, London E1 4NS, UK\\
$^{2}$Steward Observatory, University of Arizona, 933 N. Cherry Ave, Tucson, AZ 85721-0065, USA\\
$^{3}${Alien Earths Team, NASA Nexus for Exoplanet System Science, USA} \\
$^4$ Institut f\"{u}r Theoretische Astrophysik, Zentrum f\"{u}r Astronomie, Heidelberg University, Albert Ueberle Str. 2, 69120 Heidelberg, Germany \\
$^5$ Institute for Astronomy, University of Hawai'i at Manoa,2680 Woodlawn Dr., Honolulu, HI, USA \\
$^6$ Institute of Astronomy, Madingley Rd, Cambridge, CB3 0HA, UK\\
$^7$ Astrophysics Group, Imperial College London, Blackett Laboratory, Prince Consort Road, London SW7 2AZ, UK \\
$^8$ Università degli Studi di Milano, via Giovanni Celoria 16, 20133 Milano, Italy \\
$^9$ European Southern Observatory, Karl-Schwarzschild-Str. 2, 85748 Garching, Germany \\
$^{10}$ NASA Headquarters, 300 E Street SW, Washington, DC 20546, USA \\
$^{11}$Department of Physics and Astronomy, University College London, Gower Street, London, WC1E 6BT, UK \\
}

\date{Accepted XXX. Received YYY; in original form ZZZ}

\pubyear{2015}

\begin{document}
\label{firstpage}
\pagerange{\pageref{firstpage}--\pageref{lastpage}}
\maketitle

\begin{abstract}

We used the Atacama Pathfinder Experiment (APEX) telescope to search for C\,I 1-0 (492.16\,GHz) emission towards 8 proplyds in NGC 1977, which is an FUV radiation environment two orders of magnitude weaker than that irradiating the Orion Nebular Cluster (ONC) proplyds. C\,I is expected to enable us to probe the wind launching region of externally photoevaporating discs. Of the 8 targets observed, no 3$\sigma$ detections of the C\,I line were made despite reaching sensitivities deeper than the anticipated requirement for detection from prior APEX CI observations of nearby discs and models of external photoevaporation of quite massive discs. By comparing both the proplyd mass loss rates and C\,I flux constraints with a large grid of external photoevaporation simulations, we determine that the non-detections are in fact fully consistent with the models if the proplyd discs {are very low mass}. Deeper observations in C\,I and probes of the disc mass with other tracers (e.g. in the continuum and CO) {can test this}. If {such a test finds higher masses, this would imply carbon depletion in the outer disc, as has been proposed for other discs with surprisingly low C\,I fluxes}, though more massive discs would also be incompatible with models that can explain the observed mass loss rates {and C\,I non-detections}. The expected remaining lifetimes of the proplyds {are estimated to be similar to those of proplyds in the ONC} at 0.1\,Myr. Rapid destruction of discs is therefore also a feature of common, intermediate UV environments. 
\end{abstract}




\begin{figure*}
    \centering
    \includegraphics[width=2.0\columnwidth]{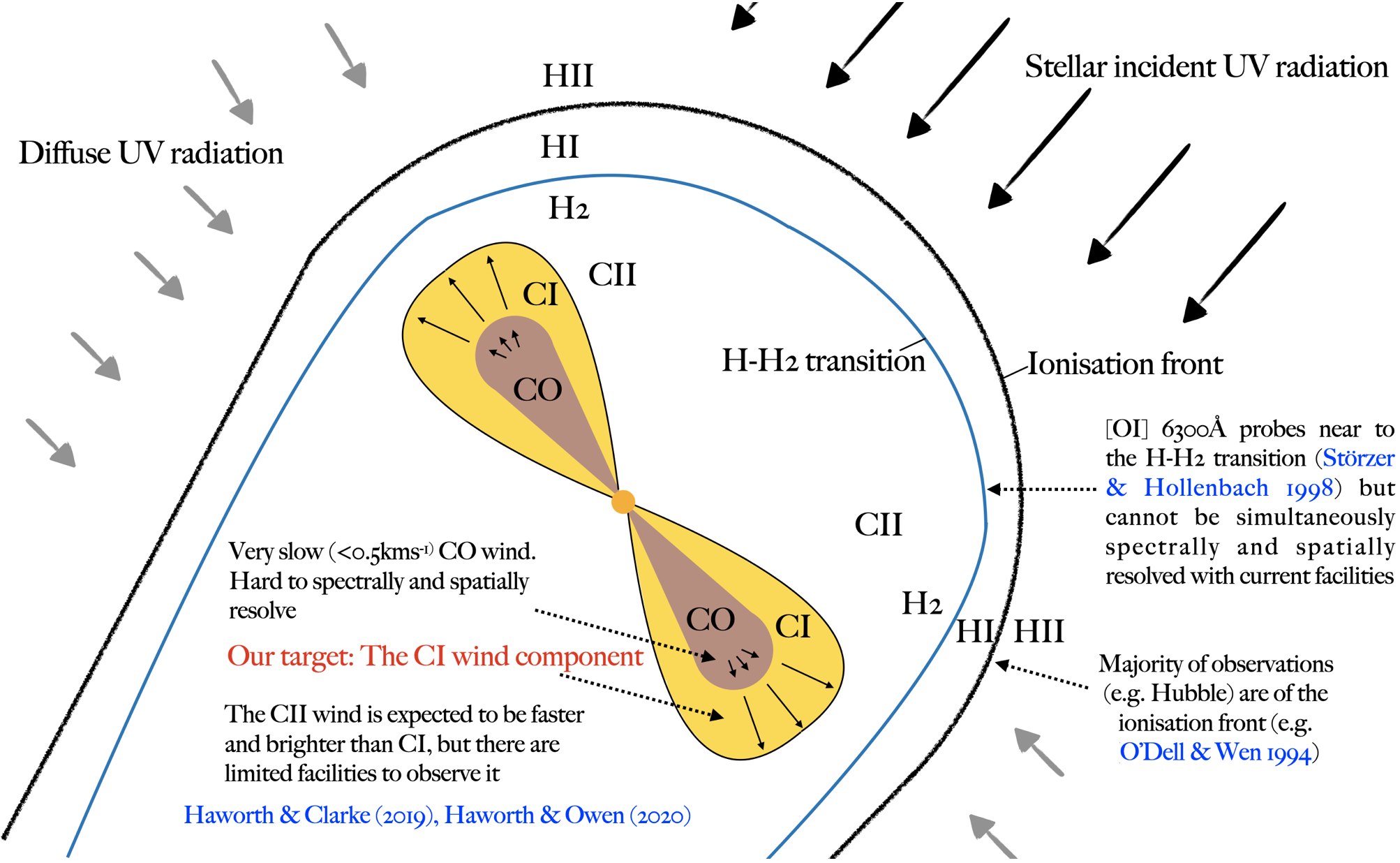}
    \caption{A cartoon of a proplyd with a breakdown of the composition, velocity and anticipated observability different wind components. The ionisation front has been observed since the mid-90's with Hubble. The H-H2 transition has been probed for a similar time indirectly via the [OI] 6300\AA\ forbidden line. However for probing the inner parts of an external photoevaporative wind we have been much more limited. CO is only present in the slowest parts of the wind before being dissociated, making it a poor tracer. CII probes a large part of the fast wind, but can only be observed spectrally with SOFIA. C\,I on the other hand should probe the faster inner part of the wind and could do so in a spectrally and spatially resolved way with ALMA. \protect\citep{1998ApJ...502L..71S}. }
    \label{fig:ProplydCartoon}
\end{figure*}

\section{Introduction}
Massive stellar clusters typically host massive stars, which dominate the production of UV photons in a cluster \citep[e.g.][]{2003ARA&A..41...57L, 2010ARA&A..48...47A}. This injection of energy into the surroundings is referred to as feedback and has a dramatic effect, driving hot bubbles that can inhibit the star formation episode that gave rise to the cluster \citep[e.g.][]{2012MNRAS.427..625W, 2015NewAR..68....1D,  2015MNRAS.450.1199D, 2015MNRAS.454.4484G, 2016MNRAS.463.3129G, 2018MNRAS.477.5422A, 2021MNRAS.501.4136A} and sculpt the pillars and bright rims that constitute some of the most stunning imagery in astronomy \citep{2006MNRAS.369..143M, 2012MNRAS.420..141E,2012A&A...546A..74D,  2015MNRAS.450.1057M}. This strong UV radiation field is also capable of heating and dispersing circumstellar protoplanetary discs, and indeed we observe that discs in high UV environments such as the central Orion Nebular Cluster (ONC) and NGC 2024 exhibit cometary outflows of material due to this ``external photoevaporation'' \citep{1994ApJ...436..194O, 1998AJ....116..293B, 1999AJ....118.2350H, 2002ApJ...566..315H, 2008AJ....136.2136R,2021MNRAS.501.3502H}. The term ``proplyd'' (originally used to describe all discs as a portmanteau of ``protoplanetary disc'') has now been adopted to describe only these cometary shaped strongly photoevaporating discs. The effect of mass loss from external photoevaporation on the disc is important, since it could limit the mass reservoir for planet formation, the timescale over which planet formation can take place and by truncating the disc could affect the viscous evolution of the entire disc and the possible radii for planet formation \citep[e.g.][]{1998AJ....116..322H, 2000ApJ...539..258R, 2017AJ....153..240A,2017MNRAS.468.1631R, 2018ApJ...860...77E, 2018MNRAS.475.5460H, 2018MNRAS.478.2700W, 2019MNRAS.490.5678C, 2020ApJ...894...74B, 2020A&A...640A..27V, 2020MNRAS.492.1279S}. 

The UV field in a cluster spans a continuum of values, and the ONC proplyds likely represent the conditions towards one extreme of that distribution in Galactic star forming regions \citep{2008ApJ...675.1361F, 2020MNRAS.491..903W}. Theoretical models predict that external photoevaporation can be important for disc evolution over a wide range of UV environments, dependent mostly upon the disc size \citep[since material in the outer part of large discs is only weakly bound to the star][]{2004ApJ...611..360A,2016MNRAS.457.3593F, 2018MNRAS.481..452H}. The UV field strength is characterised by the Habing field, unit $G_0$ which is $1.6\times10^{-3}$\,erg\,cm$^{-2}$\,s$^{-1}$ over the wavelength range 912-2400\AA\,\, \citep{1968BAN....19..421H}. The ONC represents a $\sim10^5$G$_0$ environment, whereas the solar neighbourhood is $\sim1$\,G$_0$. Recently, evidence for external photoevaporation has been found in much weaker environments than the ONC. \cite{2012ApJ...756..137B} and \cite{2016ApJ...826L..15K} discovered proplyds in close proximity of a B star in  NGC 1977 (a $\sim3\times10^3$G$_0$ environment). \cite{2017MNRAS.468L.108H} also suggest that the outer halo of the large disc IM Lup \citep{2009A&A...501..269P, 2018ApJ...865..155C} could also be due to a slow external wind in a $<10$\,G$_0$ environment (this is an exception because the disc is so extended that it is only very weakly gravitationally bound to the star). HD163296 also shows evidence for an outer wind at around 400\,au in a weak ($\leq10$\,G$_0$) environment \citep{2019Natur.574..378T, 2021ApJS..257...18T}

When a disc is strongly irradiated, mass is lost over a significant fraction of its surface. Since the irradiation is directional (approximately a planar field) this results in a cometary morphology which is referred to as a proplyd \citep[see Figure \ref{fig:ProplydCartoon} for a cartoon and][for examples]{1999AJ....118.2350H, 2016ApJ...826L..15K, 2021MNRAS.501.3502H}. In weaker UV environments, mass is only driven from the more weakly bound outer disc rather than the disc surface \citep{2019MNRAS.485.3895H}, so the cometary morphology disappears and the disc/wind is hard to distinguish from a non-evaporating disc. Furthermore the wind is typically going to be more extended than the commonly observed mm continuum, which is known to be more compact than the gas disc \citep[e.g.][]{2017A&A...605A..16F, 2019A&A...629A..79T, 2021MNRAS.507..818T, 2021A&A...649A..19S} and the commonly observed CO is also dissociated before the wind obtains an easily spectrally resolved velocity \citep[e.g.][]{2018MNRAS.481..452H, 2019MNRAS.485.3895H}. New tracers are therefore required to identify external photoevaporation in intermediate UV environments when externally photoevaporating discs do not look like proplyds. To this end, \cite{2020MNRAS.492.5030H} predicted that atomic carbon is a good kinematic probe of the inner wind. A cartoon schematic of a proplyd and the locations of various tracers is shown in Figure \ref{fig:ProplydCartoon}.

\cite{2020MNRAS.492.5030H} demonstrated that the CI 1-0 line offers unambiguous signatures of a wind in PV diagrams and can be used to diagnose components of the flow velocity, constrain the flow temperature and when coupled with models can be used to estimate the mass loss rate. The goal of this paper is to use the Atacama Pathfinder Experiment telescope (APEX) \citep[which does not have the spatial resolution to undertake all of the diagnostics proposed by][]{2020MNRAS.492.5030H} to constrain the atomic carbon line fluxes of known evaporating discs for future ALMA observations.

\cite{2016A&A...588A.108K} obtained the majority of existing C\,I observations towards discs, which were also obtained using APEX. They surveyed nearby systems, detecting C\,I 1-0 in 6 of 12 targets. They interpreted their observations with DALI astrochemistry/radiative transfer simulations \citep{2012A&A...541A..91B, 2013A&A...559A..46B}. They \citep[and][]{2012A&A...541A..91B} found that only HD 100546 was consistent with having a carbon abundance like that in the interstellar medium (ISM). All others have to be depleted in carbon by 1-2 orders of magnitude. \cite{2013ApJ...776L..38F} also inferred a deficiency of up to 2 orders of magnitude depletion in C$^{18}$O abundance which could be due to it being reacted out, freeze out or isotopolog-selective CO photodissociation \citep{2014A&A...572A..96M}. Carbon depletion has also been inferred in other systems, for example by comparing CO and HD mass estimates \citep{2016ApJ...831..167M} and in the recent MAPS ALMA large program, \citep{2021ApJS..257....1O, 2021ApJS..257....7B}. Overall, we proceed with a high uncertainty on the carbon abundance and an expectation that it is likely typically depleted. If carbon depletion is confirmed, it could also have important implications for understanding mass estimates using carbon bearing molecules. Given the low fluxes due to probable depletion, \cite{2016A&A...588A.108K} concluded that ALMA would be required over APEX for surveying C\,I in discs beyond the closest systems.

\begin{table*}
    \centering
   \begin{tabular}{lcccccccccccccccccc}
    \hline
    Target & RA & DEC &  Time on  &  RMS & Upper limit on line flux   & Mass loss rate  & Stellar  & K \\
    & (J2000) & (J2000) & source  &   (mK) & from RMS assuming   & estimate & mass &   \\
    & & &(minutes) &  & 4km/s FWHM  (K\,km\,s$^{-1}$)& ($10^{-8}M_\odot$yr$^{-1}$) &  ($M_\odot$) & (mag)  \\
    \hline
   KCFF2016--1 & 5 35 24.14 & -4 50 09.2 & 23.3  & 27.85  & $0.12$ & -- & $\sim0.2$  & 11.72$\pm0.001$  \\    
   KCFF2016--2 (V2438-Ori) & 5 35 25.50  &  -4 51 20.1 & 35.0   & 22.43  &  $9.6\times10^{-2}$ & 17.8 & $\sim0.4$  & 10.70$\pm0.001$ \\
    KCFF2016--3 & 5 35 28.81 &  -4 50 22.0 & 17.6  &  28.05  &  0.12 & 2.4  & $\lesssim 0.020$ &  16.40 $\pm0.042$ \\
    KCFF2016--4 & 5 35 28.07 & -4 50 02.1 & 58.3 &   10.81 & $4.6\times10^{-2}$ & 3.3 & $\lesssim 0.015$  & 17.77 $\pm0.149$ \\
   KCFF2016--5 & 5 35 23.38 & -4 51 18.1 & 11.6   & 33.95 & 0.14 & 4.6 & $\lesssim 0.015$ & 17.14 $\pm0.082$ \\
    KCFF2016--6 (V2390 Ori) &  5 35 22.5 & -4 52 36.5 & 34.3    & 19.03 & $8.1\times10^{-2}$  & 3.6 & $\sim0.5 $ & 10.59 $\pm0.001$   \\
    KCFF2016--7 & 5 35 23.13 & -4 48 27.7 & 17.5  & 35.8 & 0.15 & 3.1 & $\sim 0.15$   & 14.68 $\pm0.001$ \\
    The spindle (V418 Ori) & 5 35 28.69 & -4 48 16.4  & 11.7  & 26.35 &  0.11 & $\sim1$ &  $\lesssim 1.0$  &  13.48 $\pm0.003$\\
    \hline     
    \end{tabular}
    \caption{Summary of our C\,I 1-0 observations. We have no 3$\sigma$ detections, but provide the RMS noise achieved at $\Delta v = $0.48\,km\,s$^{-1}$ channel widths. The integrated line flux estimate is an upper limit assuming a peak brightness temperature of the RMS and the expected typical FWHM in the C\,I region of 4\,km\,s$^{-1}$. The mass loss rate is estimated from ionisation balance using the radius of the ionisation front and incident ionising flux. K band magnitudes are from the UKIDSS-DR9 catalog and stellar mass estimates made using pre-main sequence isochrones and mass tracks from \protect\cite{2015A&A...577A..42B}. }
    \label{tab:observations}
\end{table*}

This paper presents the results of APEX observations to gauge the C\,I brightness towards known externally photoevaporating discs in the intermediate UV environment of NGC 1977, where the main UV source is the B star 42 Ori. The goal is to lay the foundation for ALMA observations in C\,I of externally photoevaporating discs to utilise the diagnostics proposed by \cite{2020MNRAS.492.5030H}.

\section{Observations}
\label{sec:observations}

The APEX NFLASH instrument was used in service mode to observe 8 proplyds in the intermediate UV environment around the B star 42 Orionis in NGC 1977 (program ID: 108.21WZ, \textit{Probing the external photoevaporation of planet-forming discs with CI}, PI: Haworth). Our targets were the 7 proplyds identified by \cite{2016ApJ...826L..15K} which refer to as KCFF2016-1 to KCFF2016-7, and the single proplyd ``the spindle'' identified by \cite{2012ApJ...756..137B}.

Our initial proposal was to utilise the 460 and 230 receivers of NFLASH to simultaneously observe C\,I 1-0 and CO 2-1 respectively, however the installation of that dual observing capability was delayed by the covid-19 pandemic. We therefore focused on the C\,I 1-0 (492.16\,GHz) observations only. Based on 2D radiation hydrodynamic simulations and synthetic observations from \cite{2020MNRAS.492.5030H} which had similar line fluxes to the APEX C\,I observations of \cite{2016A&A...588A.108K} the anticipated peak brightness temperature in the $12.7\arcsec$ beam was 0.16\,K and so an RMS noise of 32mK (anticipated SNR of 5) was requested for a channel width of 0.5\,km\,s$^{-1}$ (hereafter when we refer to the RMS it is for this channel width unless specified otherwise). The observations themselves used wobbler switching mode.

The observations were taken on 23rd August--2nd September 2021 in good conditions (pwv 0.33-0.5\,mm), meaning that the SNR achieved was better than anticipated. However the first target, KCFF2016-2, resulted in a non-detection. We therefore adopted a strategy of observing all targets down to at least 35\,mK (at $\Delta v = 0.48\,$km\,s$^{-1}$) and then choosing any promising looking targets to go deeper on. Overall this led to 7 non-detections with RMS in the range 19.02--35.8\,mK (at $\Delta v = 0.48\,$km\,s$^{-1}$, but we also had no detections when smoothing) and a  $2.1\,\sigma$ line for one target (KCFF2016-4) at an RMS of $5.19$\,mK for $\Delta v = 1.86\,$km\,s$^{-1}$ (RMS of $10.81$\,mK for $\Delta v = 0.48\,$km\,s$^{-1}$)  which if real would have a line flux of $4.25\times10^{-2}$\,K\,km\,s$^{-1}$, peak brightness temperature of $10.9\,$mK and line full width at half maximum (line width) of 3.66\,km\,s$^{-1}$. Note that this line width is consistent with what is expected for external photoevaporation \cite[$\sim4$km\,s$^{-1}$][]{2020MNRAS.492.5030H}. Alternatively, if we assume that the possible line is just noise and use the RMS as an upper limit on the peak temperature of the line and assume a Gaussian line width of 4\,km\,s$^{-1}$ we get an upper limit on the integrated line of $2.2\times10^{-2}$K\,km\,s$^{-1}$. Since a $\sim4$\,km\,s$^{-1}$ line width is well motivated by hydrodynamic models we also assumed that value to estimate upper limits on the integrated line flux for the other targets. A summary of the observation parameters and these limits on the integrated flux using the $\Delta v=0.48$km\,s$^{-1}$ RMS values is given in table \ref{tab:observations}.

We estimate the mass loss rate for each system using the criterion of photoionisation equilibrium at the ionisation front \citep{1998ApJ...499..758J}, which relates the mass loss rate to the size of the cometary cusp of the proplyd (the radius of the ionisation front, $R_{\textrm{IF}}$) and ionising flux incident upon the proplyd
\begin{equation}
   \left(\frac{\dot{M}_w}{10^{-8}\,M_\odot\,yr^{-1}}\right) =  \left(\frac{1}{1200}\right)^{3/2}\left(\frac{R_{\textrm{IF}}}{\textrm{au}}\right)^{3/2}\left(\frac{d_{\textrm{sep}}}{\textrm{pc}}\right)^{-1}\left(\frac{\dot{N}_{ly}}{10^{45}\,s^{-1}}\right)^{1/2}. 
   \label{equn:Mdot}
\end{equation}
We adopt $\dot{N}_{ly} =2\times10^{45}$ as the ionising photon output from 42 Ori \citep{2012ApJ...756..137B} and use the projected separations $d_{\textrm{sep}}$ and ionisation front radii from Table 1 of \cite{2016ApJ...826L..15K}. 

In the discussion that follows we will compare the mass loss rates and constraints on C\,I integrated line fluxes with a subset of models from a new version of the FRIED grid of externally photoevaporating disc models \citep[][Haworth et al., in preparation]{2018MNRAS.481..452H}. The details of the modelling are given below, but we note here that the new models we consider are for discs with central stars of 0.1, 0.3 and 0.6\,M$_\odot$. We therefore require a rough constraint on the stellar mass to determine which model set each target is best compared against.

\begin{figure}
    \centering
    \includegraphics[width=\columnwidth]{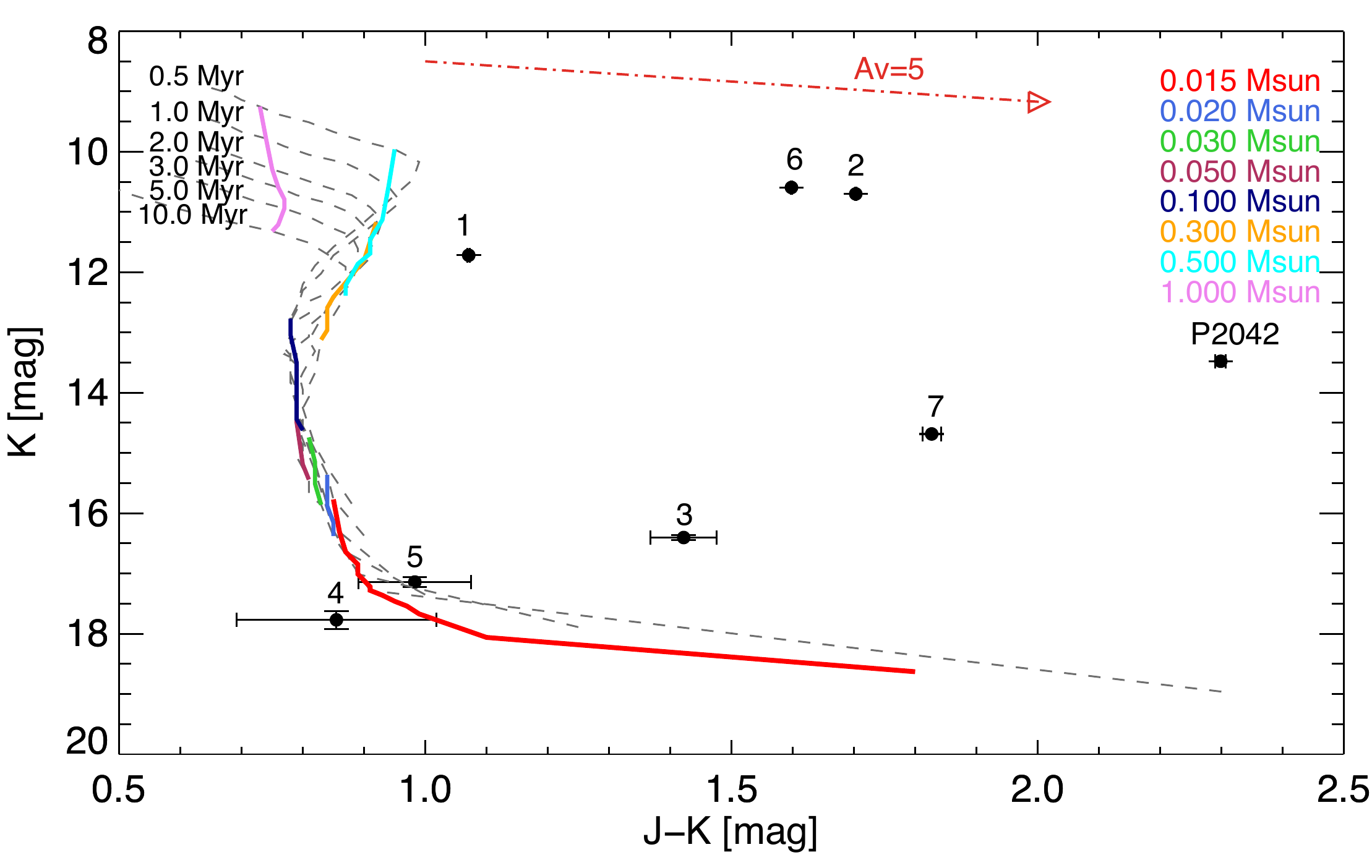}
    \caption{{A colour magnitude diagram with photometry for the known proplyds in NGC 1977. The evolutionary tracks (coloured solid lines) are \protect\cite{2015A&A...577A..42B} evolutionary models for masses from 0.015 M$_\odot$ to 1.0 M$_\odot$. The grey dashed lines are isochrones for 0.5. 1.0, 2.0, 3.0, 5.0, and 10.0 Myr. Reddening (e.g. due to the disc inclination) shifts the J-K colour to the right in this plot, with a reddening vector provided towards the top-centre of the plot. The K band magnitudes alone coupled with the evolutionary tracks allow us to make stellar mass estimates. KCFF 3, 4 and 5 are hence expected to have extremely low masses (see Table \ref{tab:observations}). } }
    \label{fig:CMD}
\end{figure}

We place constraints on the host star masses for proplyds using J,H,K photometry from the  UKIDSS-DR9 catalog {\citep{2007MNRAS.379.1599L, 2013yCat.2319....0L}} and the pre-main sequence isochrones and mass tracks from \cite{2015A&A...577A..42B} calculated for the UKIDSS J, H, K filters. We estimate ranges of the masses for the KCFF2016-3,4, and 5 using their JHK magnitudes for $A_{V}=0.5, 5.0, 20.0$\,mag using a reddening of R=5.5, which fitted better for young stars in the Trapezium cluster \citep{2021ApJ...908...49F}. {Figure \ref{fig:CMD} shows the (J-K), K colour-magnitude diagram for the NGC 1977 proplyds, a reddening vector, and \cite{2015A&A...577A..42B} pre-main sequence evolutionary tracks up to 10\,Myr. } Based on the K band magnitudes {(which are relatively insensitive to the reddening as highlighted by the vector in Figure \ref{fig:CMD})} and model predictions, we find that KCFF2016-3,4 and 5 are remarkably low mass objects $\lesssim$0.015 M$_\odot$, which correspond to brown dwarf or super planetary (free-floating planet) masses. However, {given such low masses are perhaps surprising,} in our analysis of those lowest mass systems using simulations of external photoevaporation we will use models with a stellar mass of 0.1\,M$_\odot$. This is conservative since lower mass stars have their discs unbound in external photoevaporation more easily. 

Drawing 8 random stars from a \cite{2001MNRAS.322..231K} initial mass function (IMF) yields at least 4 of 8 stars with masses $\leq0.1$\,M$_\odot$ around 57\,per cent of the time, but 3 of 8 stars with masses $\leq0.015$\,M$_\odot$ only around 2.1\,per cent of the time. So it is unlikely that we have fortuitously observed three such low mass objects simultaneously under a \cite{2001MNRAS.322..231K} IMF. We identify three possible explanations for this. First, it could be that they are simply slightly higher mass (3 of 8 stars at $\leq0.1$\,M$_\odot$ is drawn 82\,per cent of the time). Secondly, the sub-stellar IMF is not well constrained at the very low mass end \citep[down to $\leq0.03$\,M$_\odot$, ][]{2003PASP..115..763C, 2013pss5.book..115K}, so it could be that there is a change in the form of the distribution there.  Discs around low mass objects are weakly bound and hence easily externally photoevaporated, so if $\leq0.03$\,M$_\odot$ objects are more abundant they may just be easier to identify in the vicinity of strong UV sources because of the dispersing disc. If we modify the \cite{2001MNRAS.322..231K} IMF to include a component that scales as $M^{-2}$ below $\leq0.03$\,M$_\odot$ we draw 3 of 8 objects with $\leq0.015$\,M$_\odot$ 67\,per cent of the time. Proplyds could therefore provide a unique opportunity to study the extremely low mass end of the sub-stellar mass function. In support of this idea, \cite{2016MNRAS.461.1734D} infer a bimodal distribution for the IMF in the Orion Molecular Cloud with a secondary peak at 0.025\,M$_\odot$. {\cite{2020ApJ...896...80G} also found that the standard IMF under-predicts the number of very low mass objects (albeit without a bimodal distribution)}. Furthermore \cite{2021NatAs.tmp..248M} find a rich population of free floating planets towards Upper Sco. As mentioned above, it is plausible that a large underlying low mass population exists and that it is just easier to detect in strong UV environments due to the comparatively bright proplyds. Finally, a high UV environment might favourably produce low mass objects by stripping their natal star forming material \citep{2004A&A...427..299W}. 

We also note that other proplyds with very low (perhaps even planetary) masses have been discovered elsewhere, such as the $<13$\,M$_{\textrm{jup}}$ (M9.5) proplyd 133-353 in the ONC \citep{2010AJ....139..950R, 2016ApJ...833L..16F}. So although the number of low mass objects in our sample is unexpectedly high, they are not unique as proplyds.

\section{Modelling and Discussion}
To recap, our observations of 8 targets yielded no strong detections. Atomic carbon emission from these discs is therefore weaker than anticipated based on the similar observations of non-proplyds by \cite{2016A&A...588A.108K} and the models of \cite{2020MNRAS.492.5030H}. Here we discuss why that is the case, supported by radiation hydrodynamic models of externally irradiated discs and the implications this has for external photoevaporation. 

\subsection{Radiation hydrodynamic simulations and model C\,I flux estimates}
To assist in the interpretation of our observations we use 1D spherical radiation hydrodynamic simulations of external photoevaporation with the \textsc{torus-3dpdr} code \citep{2015MNRAS.454.2828B, 2019A&C....27...63H}. Using 1D models allows us to explore the anticipated line flux over a wide parameter space and has been demonstrated to give mass loss rates that are in good agreement with 2D models \cite{2019MNRAS.485.3895H}. Note that the CI flux in the 2D models is slightly higher ($\sim50$\,per cent) compared to analogous 1D simulations. 

The approach to running these 1D models is thoroughly detailed in \cite{2018MNRAS.481..452H}, where a large collection of them was generated in the FRIED public grid of mass loss rates. To summarise briefly though, they involve iteratively solving grid based hydrodynamics and photodissociation region (PDR) physics. A disc is imposed with surface density of the form 
\begin{equation}
    \Sigma(R) = \Sigma_{1\textrm{au}}\left(\frac{R}{\textrm{au}}\right)^{-1}
    \label{equn:Sigma}
\end{equation}
which is then irradiated and the steady state flow from the disc solved for along the mid-plane. Mass loss rates are estimated by assuming that the mid-plane flow applies over the whole solid angle subtended by the disc outer edge from the star assuming azimuthal symmetry \citep[see also][]{2004ApJ...611..360A, 2016MNRAS.457.3593F}. The models included here are a small subset of a new grid that is being prepared for a next generation version of FRIED ({F}UV {R}adiation {I}nduced {E}vaporation of {D}iscs, Haworth et al. in preparation). They cover stellar masses from $0.1-3$\,M$_\odot$, $\Sigma_{1au}$ from $10-10^5$\,g\,cm$^{-2}$, UV fields from $10-10^5$\,G$_0$ and disc radii from $1-500$\,au (the disc radius is that at which we fix a boundary condition to the wind). The new FRIED grid will permit selection of whether grain growth in the disc has occurred and a choice of PAH abundance. Here we only consider models where grain growth in the disc has occurred, which means less dust is entrained in any wind. We assume a dust-to-gas mass ratio in the wind of $10^{-4}$ and a cross section $\sigma_{\textrm{FUV}}=5.5\times^{-23}$\,cm$^{2}$. The PAH abundance in the outer disc is poorly constrained, but a key heating contributor, so we use a PAH-to-dust mass ratio that is half that of the interstellar medium. We do not include a detailed look at the models themselves here, but note that they have been checked for consistency against the \cite{2018MNRAS.481..452H} calculations. 

Because the PDR chemistry is solved as part of the dynamical model it is trivial to estimate the C\,I line flux using the PDR abundances and temperature distribution. We assume that the CI emission predominantly comes from the wind, so ignore any emission that would have come from within the disc itself. {We do this because within the disc itself the PDR chemical network is not a complete description of the chemistry. Taking this approach means that our estimates of the CI emission are conservatively low. Note that in our 2D models C\,I emission from anywhere is permitted and gives fluxes to within around a factor 2 of the 1D models.} To estimate the C\,I line flux from these 1D models we use the PDR abundance of atomic carbon, but assume local thermodynamic equilibrium to solve the level populations and assume that the line is optically thin. 

Specifically, the emission coefficient in any given cell of the grid is 
\begin{equation}
    j_\nu = \frac{1}{4\pi}n_uA_{ul}h\nu
\end{equation}
where $n_u$ is the number of carbon atoms in the upper state of the transition (so dependent upon the spatially varying PDR calculation of the carbon abundance), $A_{ul}$ is the Einstein A coefficient of the transition from $u$ to $l$ and $\nu$ is the transition frequency. Since we are assuming the line is optically thin, the total emergent intensity is obtained simply by integrating the emission coefficient from the disc outer edge, through the wind to the outer edge of the grid $I_\nu = \int j_\nu dl$. This intensity is then converted to a mock APEX integrated flux in K kms$^{-1}$ by assuming that the disc is not spatially or spectrally resolved and that the emission comes from a characteristic surface that is a function of the disc size ($\pi R_d^2$). That is, if the flux at distance $D$ in W\,m$^{-2}$ is
\begin{equation}
    F_0 = \frac{\pi R_d^2}{D^2}\int j_\nu dl    
\end{equation}
The unresolved flux in K\,km\,s$^{-1}$ is then
\begin{equation}
    F =1.149\times10^{29} F_0    \left(\frac{\theta_{\textsc{apex}}}{\textrm{arsec}}\right)^{-2} \left(\frac{\nu}{\textrm{GHz}}\right)^{-3}
\end{equation}
where $\theta_{\textsc{apex}}$ is the APEX 12.7 arcsecond beam. 

The rationale for using these models is to determine the broad scaling of the C\,I flux with UV field strength and the star-disc parameters. Furthermore, in the case of the proplyds we also have observational mass loss estimates (from equation \ref{equn:Mdot}) so we can compare models and observations using trends in both the C\,I flux and mass loss rates. The goal is not to use the models to provide detailed modelling of any given system.

\subsection{Why is the C\,I emission fainter than expected?}
There are various possible reasons that the C\,I emission could be fainter than we anticipated based on the \cite{2016A&A...588A.108K} observations and the small sample of 2D models from \cite{2020MNRAS.492.5030H}. These possibilites include
\begin{enumerate}
    \item If the NGC 1977 proplyds are very compact
    \item If the NGC 1977 proplyds are very low mass
    \item If there is depletion of carbon from the gas phase in the outer disc (e.g. due to freeze out and radial drift). 
\end{enumerate}
Point (iii) is interesting both because carbon depletion was concluded to be possibly important to explain the fluxes observed by \cite{2016A&A...588A.108K} and because C\,I and C\,II emission lines are important coolants in the inner wind of irradiated discs, so depletion of carbon could conceivably lead to higher temperatures and hence higher mass loss rates. {Carbon depletion was also found to lead to higher internal photoevaporation rates by \cite{2019MNRAS.490.5596W}, though in that case it is due to the fact that carbon is a significant opacity source to the X-rays.} {Note that evidence for carbon depletion has come from CO observations towards discs, where the discrepancy cannot be fully accounted for by invoking freeze out \citep[e.g.][]{2016ApJ...828...46A, 2017ApJ...844...99L}}. 

However, first we have to assess whether the low C\,I flux could simply be a result of the proplyds being heavilty truncated or depleted of mass, which is known to have happened for proplyds in the core of the ONC and in NGC 2024 \citep[e.g.][]{2018ApJ...860...77E, 2020A&A...640A..27V, 2020ApJ...894...74B}. To this end we exploited the fact that each of our models provides both a C\,I flux and mass loss rate, both of which are constrained for the NGC 1977 proplyds. Furthermore, the mass loss rate in the models is somewhat sensitive to the host star mass, which in practice is uncertain for each target but does have some constraint. We can therefore determine whether the combined mass loss rates and C\,I fluxes are at least consistent compared to the models. 

\begin{figure*}
    \centering
    \vspace{-0.2cm}
    \includegraphics[width=1.74\columnwidth]{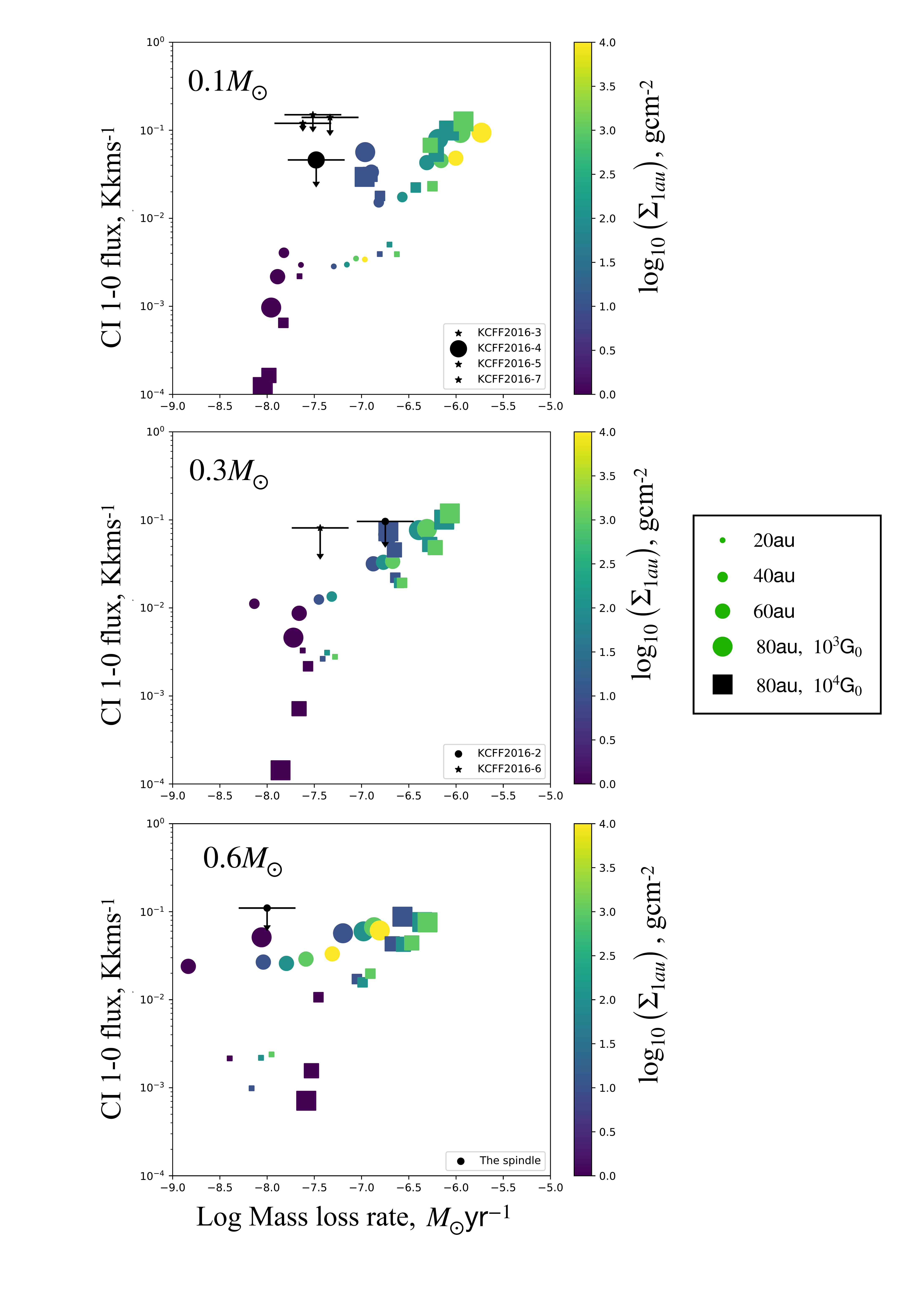}
    \vspace{-0.2cm}
    \caption{{The coloured points are next generation FRIED models in C\,I flux--mass loss rate space. Circles and squares are for models in a 10$^3$ and 10$^4$G$_0$ environment respectively and the symbol sizes represent different disc radii. The colourscale represents the surface density at 1au (and hence disc mass), where the surface density profile is given by equation \ref{equn:Sigma}. The black points are the observed CI flux constraints based on the $\Delta v=0.48$km\,s$^{-1}$ RMS values and mass loss rates for NGC 1977 proplyds. KCFF2016-4 has a radius constraint which is reflected in the symbol size.  }}
    \label{fig:keyplot}
\end{figure*}

\subsubsection{Models in C\,I flux -- Mass loss rate space}
The information we have on the proplyds in NGC 1977 is primarily from their mass loss rate (based on the size of the ionisation front cusp, Equation \ref{equn:Mdot}) and the new C\,I flux constraints. In Figure \ref{fig:keyplot} we therefore compare our models and the observations in C\,I flux -- mass loss rate space. The mass loss rate is sensitive to the stellar mass, disc mass and disc radius, so is a convenient way of encapsulating all three parameters.

The symbol sizes of the simulation points in Figure \ref{fig:keyplot} represent the disc radius, with the smallest points being 20\,au and increasing in 20au intervals up to 80\,au. The circles and squares are models in a $10^3$ and $10^4$\,G$_0$ environment respectively \citep[the anticipated typical UV field that the NGC 1977 proplyds are exposed to is $\sim3000$G$_0$][]{2016ApJ...826L..15K}. The colour scale represents the surface density normalization at 1\,au ($\Sigma_{1au}$ from equation \ref{equn:Sigma}). For reference, the disc mass would be $M_d = 2\pi R_d \Sigma_{1\textrm{au}} \times (1\textrm{au})$, so large yellow points represent the most massive discs and small black points the least massive discs. From top to bottom the panels are for discs orbiting 0.1, 0.3 and 0.6\,M$_\odot$ stars respectively. The black points/limits represent the NGC 1977 proplyd data.

Both the C\,I 1-0 flux and mass loss rate generally increase with increasing disc mass and radius, with brightest proplyds in C\,I undergoing the highest mass loss (though at low stellar masses the C\,I 1-0 integrated flux does remain approximately constant above $\sim10^{-6}$\,M$_\odot$\,yr$^{-1}$). For the lowest mass discs there can be an exception, with larger discs giving lower C\,I fluxes because where the disc is truncated the surface density is so low. This is a known minor issue with the approach taken by FRIED, since in reality in such a scenario the radiation penetrates deeper into the disc and the wind would end up launched from smaller radii \citep{2021MNRAS.508.2493O}. This turns out not to be an issue in practice when studying disc evolution because the outer parts of such a disc would quickly be lost to the wind.

\subsubsection{Observational constraints on NGC 1977 proplyds in C\,I flux -- Mass loss rate space}
The black upper limits in Figure \ref{fig:keyplot} represent our observations. In each panel we have only included the observed points where the host star mass is closest to the model stellar mass (where for the lowest mass objects, which could be brown dwarfs or even planetary mass we have assumed 0.1\,M$_\odot$, see section \ref{sec:observations} for a discussion on this) and have further only included points where there is an estimate of the mass loss rate. The mass loss rate errors assume a factor of 50\,per cent uncertainty in the true separation of the proplyd from 42 Ori (they are in close projected separation and are the lowest UV environment proplyds known, so they can't get much more distant than the projected separation with the mass loss rate dropping which would result in loss of proplyd morphology). The point for KCFF2016-4 is scaled in size to represent the upper limit radius constraint on the disc by \cite{2016ApJ...826L..15K}.

The key message from overlaying these upper limits on Figure \ref{fig:keyplot} is that {\textit{for the combination of proplyd mass loss rates and sensitivities achieved, the models are entirely consistent with getting only non-detections}}. 

This conclusion could only be realised by combining mass loss rates with the C\,I flux estimate and the large parameter space of models that the 1D calculations of FRIED (and its upcoming successor) enable. In particular, the drop off in flux at very low disc masses was not explored in the prior 2D models used to motivate the observational requirements (recall also that 2D models predict marginally higher C\,I fluxes, by up to around 50\,per cent).

Again using Figure \ref{fig:keyplot}, further interpretation for the lack of detection is that it is due to discs being extremely heavily depleted, especially for the low ($\leq0.4$\,M$_\odot$) stellar mass proplyds. KCFF2016--2 is the only target possibly consistent with a more massive disc, with the models supporting a possible $\Sigma_{1\textrm{au}}$ up to $10^3$\,g\,cm$^{-2}$.

Looking at the mass loss rate and flux, KCFF2016--3, 4, 5, 6 and 7 are all degenerate in disc radius, but are consistent almost exclusively only with the models that have $\Sigma_{1au}=1$\,g\,cm$^{-2}$. Now it is likely that in such an extreme instance of disc dispersal the disc deviates from a simple power law surface density profile, but for illustrative purposes if we are generous and use $\Sigma_{1au}=10$\,g\,cm$^{-2}$ and $R_d=80$\,au in an $R^{-1}$ surface density profile the resulting mass is only 0.45M$_{\textrm{Jup}}$ (Jupiter masses). With corresponding typical mass loss rates of $\sim4\times10^{-8}$\,M$_\odot$\,yr$^{-1}$ that would give a naive depletion timescale $M_{\textrm{disc}}/\dot{M}_{\textrm{wind}}$ of around 10\,kyr. Even with the generous disc masses and radii used in this estimate, these low mass stars (or brown dwarfs/massive planets) are expected to have discs that {could quickly be completely dispersed. It is important to note that the inferred wind depletion time-scale of a disc in a `proplyd-state' does not reflect the true life-time of the disc if the disc leaves the region of strong irradiation within a comparable time-scale, or if the disc truncates to a radius where external photoevaporation is no longer effective. }

\begin{figure}
    \vspace{-0.4cm}
    \includegraphics[width=\columnwidth]{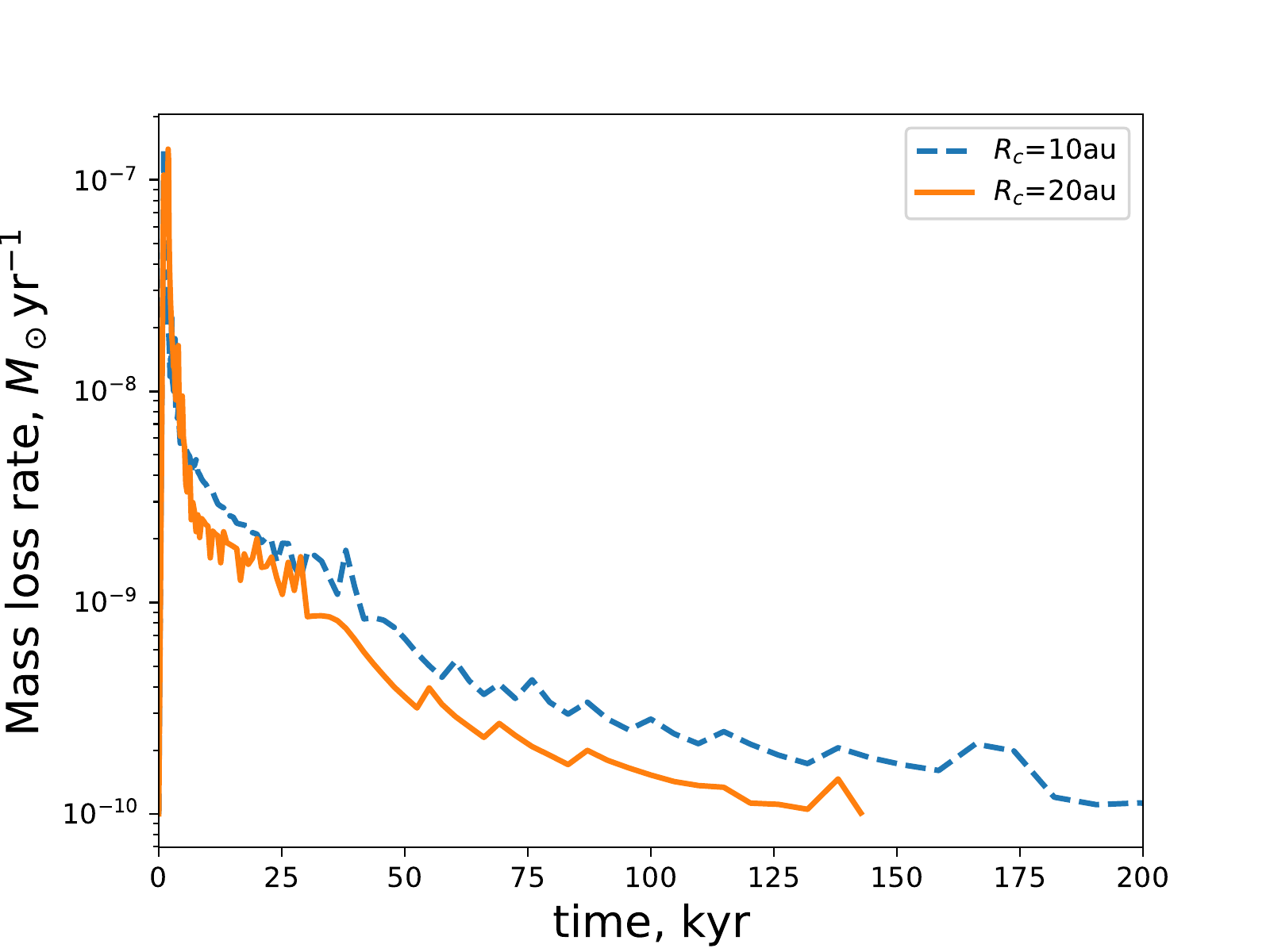}
    \vspace{-0.cm}    
    \includegraphics[width=\columnwidth]{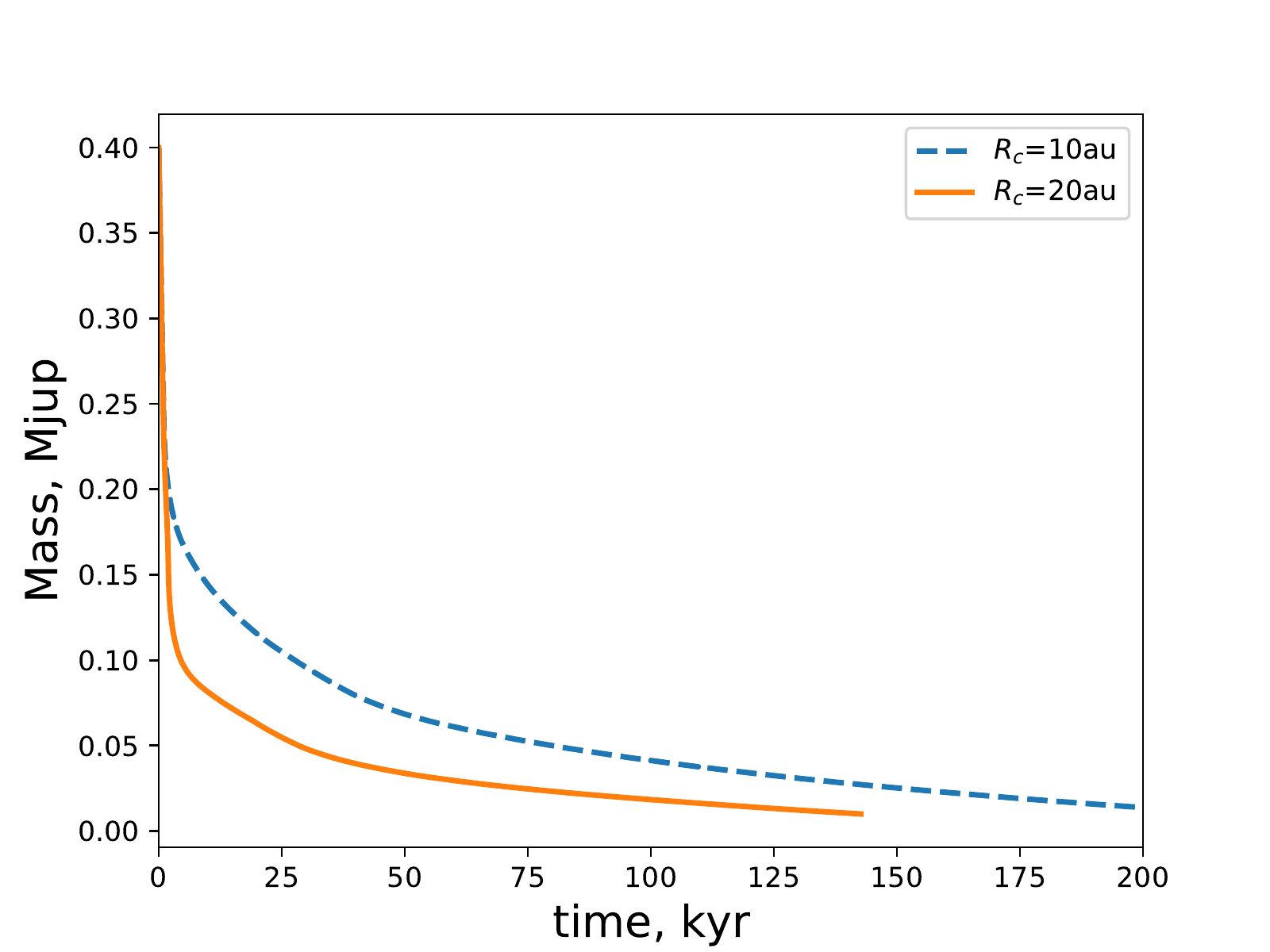}
    \vspace{-0.cm}
    \includegraphics[width=\columnwidth]{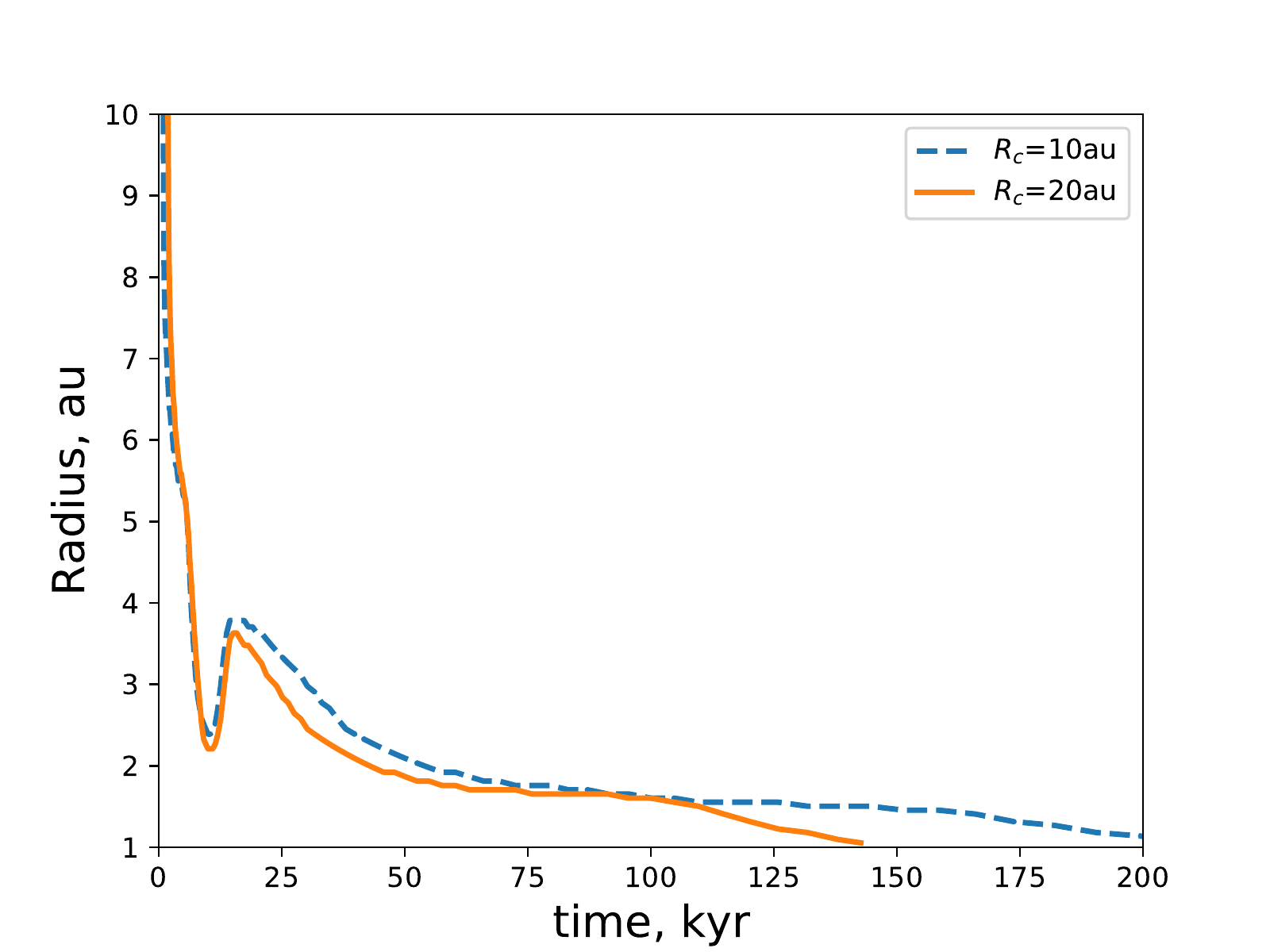}    
    \vspace{-0.3cm}
    \caption{The evolution of the mass loss rate, disc mass and radius for a disc with properties similar to those expected for the low stellar mass proplyds of NGC 1977. }
    \label{fig:QiaoModel}
\end{figure}

To get a slightly more detailed picture of the subsequent evolution we also ran a disc evolutionary model using the code of \cite{2017MNRAS.469.3994B, 2020MNRAS.492.1279S}. This calculates the disc viscous evolution including external photoevaporation by interpolating over the FRIED grid, with disc initial conditions based on a \cite{1974MNRAS.168..603L} surface density profile 
\begin{equation}
    \Sigma = \Sigma_{0}\left(\frac{R}{R_C}\right)^{-1}\exp{\left(-\frac{R}{R_C}\right)}
\end{equation}
where $R_C$ is some characteristic radius in the disc.
Based on the observations of the low stellar mass proplyds, we assume a disc mass of 0.4\,$M_{\textrm{Jup}}$, a stellar mass of 0.1\,M$_\odot$ and $R_c = 10$ and 20\,au. We use a viscous $\alpha$ of $10^{-3}$, though note that the lifetime of the disc will be sensitive to this value \citep{2020MNRAS.497L..40W}. The evolution of the mass loss rate, disc mass and disc radius is given in Figure \ref{fig:QiaoModel}, with each model terminating when the mass loss rate is $<10^{-10}\,$M$_\odot$\,yr$^{-1}$. With the caveat that at such low disc masses the surface density profile may well differ, it does provide a more detailed demonstration of the fact that these low stellar mass proplyds are on the verge of being completely dispersed on a timescale of order {0.05-0.2\,}Myr. The lifetime estimate is longer than that using the current observed mass loss rate since the mass loss rate decreases substantially with the disc radius. Note that an increase in disc radius occurs when the rate of viscous spreading at the disc outer edge is faster than the rate of external photoevaporation \citep{2007MNRAS.376.1350C}. 

{Although these NGC 1977 proplyd lifetimes are short, they are not unusually short for proplyds. The ONC proplyd lifetimes are typically expected to be about 0.1\,Myr \citep[e.g.][]{1999AJ....118.2350H}. This led to the well known ``proplyd lifetime problem'', the issue being that if proplyds only survive for a short time, we should not catch them in the act of undergoing external photoevaporation. \cite{2019MNRAS.490.5478W} demonstrated that this issue can be circumvented by ongoing migration of discs into the higher UV part of the cluster and/or the sequential exposure of embedded discs to the UV field as the star forming cloud is dispersed. Fang et al. (in preparation) will publish a census of YSOs in NGC 1977 that shows there are indeed many that are at wider separations than the proplyds and/or embedded at the periphery of the expanding radiation feedback driven bubble which could replenish the proplyd population over time, providing a resolution to the proplyd lifetime problem in NGC 1977 in the manner proposed by \cite{2019MNRAS.490.5478W}.}

{The NGC 1977 proplyds are of particular interest because they are   unambiguous examples of external photoevaporation (i.e. they are proplyds) in the weakest UV environment known to date. Their $\sim3000$G$_0$ environment is around two orders of magnitude weaker than in the core of the ONC. The implications of the low masses and hence typical proplyd lifetimes suggested by Figure \ref{fig:keyplot} would be that even in an intermediate UV environment \citep[and perhaps the most common kind of UV environment][]{2008ApJ...675.1361F, 2020MNRAS.491..903W} low mass stars are severely depleted, and rapidly, by external photoevaporation. Independent mass constraints (e.g. in the continuum, see below) will confirm this.  }

In line with theoretical expectations, the higher mass stars (centre/lower panel) which would have more bound discs are consistent with higher disc masses and brighter discs. We will discuss further in section \ref{sec:nextALMA} that these systems \citep[KCFF2016-2 and the spindle][]{2012ApJ...756..137B} should offer the best chance for detection in deeper observations.

\subsection{Next steps with ALMA/APEX }
\label{sec:nextALMA}
These APEX observations (which were undertaken in excellent conditions) and modelling have demonstrated that the higher sensitivity of ALMA is necessary to obtain detections of the C\,I 1-0 line and that KCFF2016-2 and the Spindle \citep{2012ApJ...756..137B} are the most suitable targets for deeper observations to actually obtain detections. KCFF2016-2 requires a  line flux constraint of $\sim2\times10^{-2}$\,K\,km\,s$^{-1}$ (peak of $\sim5$\,mK, requiring around 1.5 hours on the ALMA 7m array for a 3$\sigma$ detection) to rule out the models and hence require carbon depletion to explain any non-detection. They also provide expectations for the required sensitivity to obtain detections for the low mass proplyds (more like $10^{-3}$\,K\,km\,s$^{-1}$, peak of 0.23\,mK), however achieving such a sensitivity requires unrealistic integration times.

The low mass object KCFF2016-4 does have a tentative 2\,$\sigma$ line in our APEX data of the expected width and LSR velocity which, if real, would be inconsistent with the models in the upper panel of Figure \ref{fig:keyplot}. Our models predict practical deeper integration would still result in a non-detection, so confirming (or otherwise) the tentative detection would also provide a key test of the models. A prudent next step would therefore be deeper integration of C\,I 1-0 for KCFF2016-2 and KCFF2016-4. 

In addition to going deeper with C\,I 1-0, in the case of KCFF2016-2, 3, 4,5, 6 and 7 the disc mass is anticipated to be so low that \textit{any} constraints on the disc gas/dust mass and radius would be extremely valuable for determining how the final phase of disc clearing around low mass stars in common UV environments proceeds {and for confirming the implied short proplyd lifetimes (like those of proplyds in the ONC) in NGC 1977}. {Even though CO and the dust continuum do not provide the most reliable measure of the disc mass, e.g. compared to HD \citep[e.g.][]{2020A&A...634A..88K}, the constraints they could quite easily place on the dust/gas masses would still be very valuable. }

\begin{figure}
    \hspace{-0.3cm}
    \includegraphics[width=1.1\columnwidth]{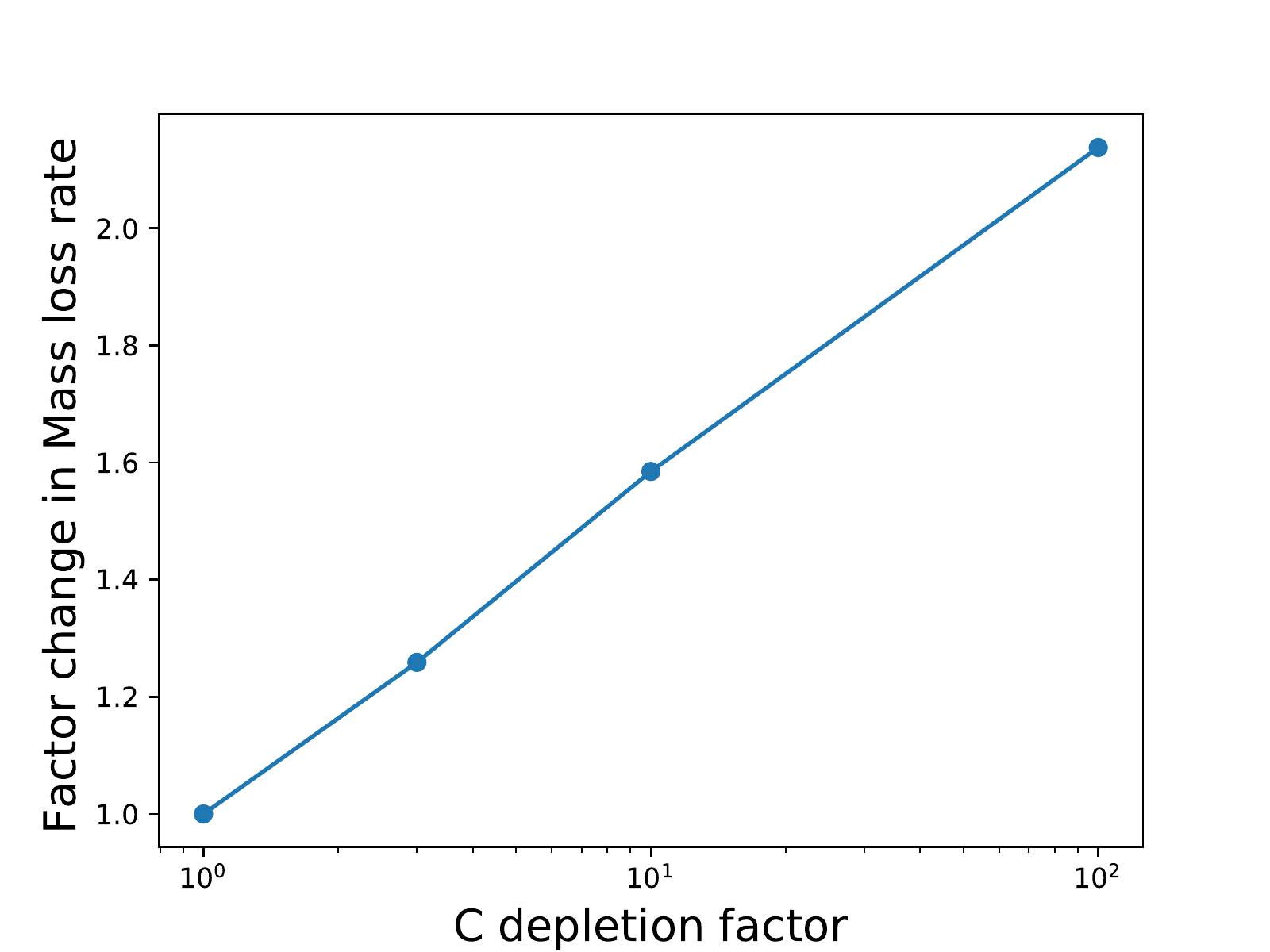}
    \vspace{-0.3cm}
    \caption{Change in external photoevaporative mass loss rate as a function of degree of carbon depletion. The CI line flux from the wind scales linearly with the C depletion.  }
    \label{fig:freezeout}
\end{figure}
\subsection{What would the consequences of carbon depletion be?}
Our models suggest that the non-detections for NGC 1977 proplyds in this paper are simply due to their discs being extremely depleted in mass. However, in the event that future observations demonstrate that this interpretation is wrong (e.g. by mass estimates from other tracers, there are currently no other constraints on the mass of the NGC 1977 proplyds) depletion of the carbon itself from the gas phase could provide an alternative explanation. Carbon depletion by, for example freeze out, was already proposed by 
\cite{2016A&A...588A.108K} to explain the lower than expected fluxes for massive, extended, isolated discs. 

To provide an initial assessment of the impact of carbon depletion on the line flux and mass loss rate we ran a small number of additional 1D external photoevaporation models with the global carbon abundance depleted by a factor 3, 10 and 100 {\citep[factors consistent with low CO versus HD mass estimates, e.g.][]{2017ASSL..445....1B}}. The models chosen were 100\,au discs with $\Sigma_{1\textrm{au}}=10^3$\,g\,cm$^{-2}$ around a solar mass star. The impact on the estimated line flux is a linear scaling with the degree of carbon depletion. That is a factor 100 decrease in the carbon abundance gives a factor 100 weaker integrated line flux. Atomic and ionised carbon lines are important coolants in external photoevaporation, so their removal also influences the mass loss rate, but as we illustrate in Figure \ref{fig:freezeout} this is not by a large factor. A factor 100 decrease in the line flux corresponds to just over a factor 2 increase in the mass loss rate. Note that \cite{2019MNRAS.490.5596W} also found an enhancement in internal photoevaporative mass loss rates due to carbon depletion, {however in that case it is due to the fact that carbon is a significant opacity source to the X-rays that drive internal photoevaporation}.

\section{Summary and conclusions}
We observed 8 known proplyds in the vicinity of a B star in NGC 1977 in C\,I emission using APEX. Our goal was to constrain the line flux to inform future ALMA observations that would enable us to probe and characterise the inner part of the externally driven wind. {In particular, C\,I has been proposed as a possible identifier and diagnostic of external winds when external photoevaporation is less extreme and so does not lead to a cometary proplyd}. {Our objective was to test its utility using proplyds in an intermediate UV environment}. The line fluxes were weaker than anticipated, resulting in no clear detections. However, the combination of C\,I flux and mass loss rate constraints with large grids of external photoevaporation models provides a useful tool for interpreting proplyds. We draw the following main conclusions from this work. \\

1) Comparing our NGC 1977 proplyd C\,I 1-0 upper limits and mass loss rates with a wide parameter space of simulations of external photoevaporation we find that the non detections are entirely consistent with the models. The fluxes being weaker than expected are simply due to the proplyds being severely depleted of mass, particularly for the proplyds associated with stars $\lesssim0.1$M$_\odot$. \\

2) The anticipated remaining lifetime of these low mass discs due to external photoevaporation is $\sim0.1\,$Myr, {so on a timescale comparable to the expected remaining lifetime of proplyds in the ONC}.  NGC 1977 is a UV environment two orders of magnitude weaker than in the ONC, excited only by a B type star, but still exhibits {the so-called proplyd lifetime problem}.   NGC 1977 does have a more extended YSO population that can circumvent this though by replenishing the photoevaporating population continuously \citep[][Fang et al. in preparation]{2019MNRAS.490.5478W}.  \\

3) The stellar masses of the proplyds are intriguingly low, with 3 out of 8  $\leq0.015$\,M$_\odot$, placing them in the low mass end of the brown dwarf or even the super-planetary regime. This is not consistent with random sampling from the IMF. Rather we suggest that either the IMF changes at the extreme low mass end of the sub-stellar regime ($<0.03$M$_\odot$) and that those objects are only easily identified when externally photoevaporated, or that radiation driven dispersal of the natal gas has favourably produced very low mass objects. {These possibilities are not new ideas, with proplyds around very low mass stars having been observed elsewhere \citep{2010AJ....139..950R, 2016ApJ...833L..16F} and with the suggestion of a bimodal IMF \citep{2016MNRAS.461.1734D}, or at least an underprediction of the standard IMF \citep{2020ApJ...896...80G} in Orion for low mass stars,  consistent with our low mass objects.  }\\

4) {Further measurements (e.g in the continuum or CO) will confirm that the low C\,I fluxes and observationally inferred mass loss rates do indeed result from low mass discs.} In the event that {such observations suggest comparatively massive discs} the explanation for lower than anticipated C\,I fluxes would most likely be carbon depletion, for example by freeze out as suggested by \cite{2016A&A...588A.108K} {to explain the low  C\,I flux in} observations towards discs in low UV environments. We made a first assessment of the impact of C depletion {upon external photoevaporation}, finding it could readily reduce the line flux, with only a small corresponding increase in the mass loss rate due to the reduced cooling by carbon emission lines. {However, our expectation of a low disc mass is based both on the C\,I flux and mass loss rate, so if the the discs are observed to be relatively massive this would also come into conflict with the models since they predict that the observed mass loss rates can only stem from low mass discs (with the exception of KCFF2016-2, which could be higher mass according to the models).}    \\

Ultimately, NGC 1977 is a fascinating laboratory for understanding external photoevaporation in an environment that is far from extreme and may even be typical. Further observations of the YSO population and winds in the region is vitally important for understanding disc evolution and planet formation in the context of their stellar clusters.

\section*{Acknowledgements}
{We thank the referee for their time and effort in providing valuable comments on the manuscript.}

This publication is based on data acquired with the Atacama Pathfinder Experiment
(APEX) under programme ID 108.21WZ. APEX is a collaboration between the
Max-Planck-Institut fur Radioastronomie, the European Southern Observatory,
and the Onsala Space Observatory. The authors are extremely grateful to the Carlos De Breuck and Michele Ginolfi who undertook the observations at APEX and provided guidance and clear and rapid communication during the observing.

TJH is funded by a Royal Society Dorothy Hodgkin Fellowship. 

CJC acknowledge support
from the STFC consolidated grant ST/S000623/1. This work has
also been supported by the European Union’s Horizon 2020 research
and innovation programme under the Marie Sklodowska-
Curie grant agreement No 823823 (DUSTBUSTERS).

JEO is supported by a Royal Society University Research Fellowship. This project has received funding from the European Research Council (ERC) under the European Union’s Horizon 2020 research and innovation programme (Grant agree- ment No. 853022, ERC-STG-2019 grant, PEVAP).

This work utilised the DiRAC Data Intensive service at Leicester, operated by the University of Leicester IT Services, which forms part of the STFC DiRAC HPC Facility (www.dirac.ac.uk). The equipment was funded by BEIS capital funding via STFC capital grants ST/K000373/1 and ST/R002363/1 and STFC DiRAC Operations grant ST/R001014/1. DiRAC is part of the National e-Infrastructure.

This research also utilised Queen Mary's Apocrita HPC facility, supported by QMUL Research-IT (http://doi.org/10.5281/zenodo.438045).

\section*{Data Availability}
The proprietory period on the data in this paper ends on 6th September 2022, at which time it will become publicly available through the ESO Science Archive Facility. Before then, the authors are happy to share the data on request.  A publication presenting the next generation FRIED models is currently being prepared, at which point the grid will be made publicly available.



\bibliographystyle{mnras}
\bibliography{example} 




\bsp	
\label{lastpage}
\end{document}